\documentstyle[12pt,aasms4]{article}
\begin{document}
\title{Stochastic Acceleration and Non-Thermal
Radiation in Clusters of Galaxies}
\author{Pasquale Blasi}
\affil{NASA/Fermilab Astrophysics Group, Fermi National Accelerator 
Laboratory, Box 500, Batavia, IL 60510-0500}

%\maketitle

\begin{abstract}
We calculate the distribution of electrons in 
clusters of galaxies, resulting from thermalization processes
in the presence of stochastic acceleration due to plasma waves. We show that 
the electron distribution can deviate from a 
Maxwell-Boltzmann distribution, due to the 
effect of the stochastic energy gain, provided waves 
can be sustained against damping. The non-thermal 
tail of electrons can generate as bremsstrahlung emission
a flux of hard X-rays compatible with the ones recently detected 
in some clusters of galaxies. 
\end{abstract}
{\it Subject Headings:} galaxies: clusters: general --- X rays: 
theory --- acceleration of particles

%\twocolumn

\section{Introduction}

The recent detection of hard X-rays (Fusco-Femiano et al. 1999\markcite{fusco};
Kaastra et al. 1999\markcite{kaastra}) and ultraviolet (UV)
radiation (e.g. Lieu et al. 1996a, 1996b\markcite{lieu1,lieu2}) 
from the Coma cluster 
and other clusters of galaxies
have raised important questions about our understanding of the
cosmic ray (CR) energy content of these large scale structures.
In the case of the Coma cluster, the detection of both
radio radiation and hard X-rays, together with the
standard interpretation based on a synchrotron plus inverse compton
scattering (ICS) model implies unusually large
CR energy densities, comparable with the equipartition value
(Lieu, et al. 1999)\markcite{lieu} and, at the same time,  
magnetic fields ($B\sim 0.1\mu G$) appreciably smaller than the ones inferred
from Faraday rotation measurements (Kim et al. 1990; Feretti et al. 1995
\markcite{kim,feretti}).

Although the condition of equipartition of CRs with thermal
energy in clusters is certainly appealing from many points of
view, it also encounters several problems: first, as
shown by Berezinsky, Blasi \& Ptuskin (1997)\markcite{bbp}, 
ordinary sources of CRs in clusters
(e.g. normal galaxies, radio galaxies and accretion shocks) can
provide only a small fraction of the equipartition energy density.
Second, gamma ray observations put strong limits on the CR energy 
density $\epsilon_{CR}$: As shown by
Blasi (1999)\markcite{blasi}, some models of CR injection that 
imply equipartition
are already ruled out by the present EGRET gamma ray observations. 
In the same paper 
it was also proposed that current and future experiments working in the
$10-1000$ GeV energy range could impose considerably better 
limits on $\epsilon_{CR}$.

Since the requirement for CR equipartition descends directly
from the assumption of ICS as the source of the hard
X-ray emission, it is natural to look for
alternative interpretations.
Ensslin, Lieu \& Biermann (1999)\markcite{ens1} showed that,
assuming a non-thermal tail in the thermal distribution of the
intracluster electrons, X-ray
observations above $\sim 20$ keV could be explained as
bremsstrahlung emission (see also Sarazin (1999)\markcite{sar99}).
The natural mechanism responsible for
this tail is stochastic acceleration of low energy electrons due
to plasma waves.
We carry out here a detailed calculation, solving numerically
a Fokker-Planck equation to derive the global (thermal plus non-thermal)
electron distribution, resulting from the superposition of thermalization
processes (mainly Coulomb scattering), acceleration and radiative energy
losses (synchrotron emission and ICS).
The electron distribution is shown to be substantially changed by
the systematic stochastic energy gain, so that hard X-rays 
can be produced by bremsstrahlung with a non-thermal spectrum. 

The paper is planned as follows: in section 2 we discuss the
stochastic acceleration, while in section 3 we outline the 
mathematical formalism used to describe acceleration and thermalization 
processes. In section 4 we apply the calculation to the case
of the Coma cluster and give our conclusions in section 5.

\section{Stochastic Acceleration}

Waves excited in a magnetized plasma can carry both an electric
field parallel and perpendicular to the direction of the magnetic
field $\vec B$. This electric field affects the dynamics of the
charged particles in the medium in several ways. In particular,
particles can be resonantly accelerated by waves when the
following resonance condition is fulfilled:
\begin{equation}
\omega - k_\parallel v_\parallel - \frac{l\Omega}{\gamma} = 0,
\label{eq:res}
\end{equation}
where $\omega$ is the frequency of the wave, $k_\parallel$ is the
parallel wavenumber, $v_\parallel$ is the component of the
velocity of the particle parallel to the direction of the magnetic
field $\vec B$, $\Omega=|q|B/mc$ is the non relativistic gyrofrequency
of the particle with electric charge $q$ and mass $m$, and finally
$\gamma$ is the Lorentz factor of the particle. The quantum number
$\ell$ is zero for resonance with the parallel electric field and can
be $\pm 1,~\pm 2,~...$ for resonances with the perpendicular
electric field.

In the present calculation we limit our attention 
to electrons as accelerated particles
(therefore $\Omega=\Omega_e=|q|B/m_e c$ in eq. (\ref{eq:res}))
and we consider fast modes as accelerating waves. These modes
populate the range of frequencies below the
proton cyclotron frequency $\Omega_p=|q|B/m_p c$, and, far from
this frequency their dispersion relation can be written as $\omega=
v_A k_\parallel$, where $v_A$ is the Alfv\`en speed.
Using now this dispersion relation and $\ell=-1$ in eq. (\ref{eq:res}),
we easily obtain the minimum momentum of the particles that can
resonate with the waves, $p_{min}\approx (m_p/m_e)\beta_A/\mu$. Here $p$
is the electron momentum in units of $m_e c$, $\beta_A=v_A/c$, 
and $\mu$ is the cosine of the pitch angle. For parameters typical
of clusters of galaxies ($B\sim 1\mu G$, average gas density $n\sim 4\times
10^{-4}cm^{-3}$), we obtain $p\gtrsim 0.5/\mu$. Particles with smaller
momenta can be accelerated by whistlers, that populate the region
of frequencies between the proton and electron cyclotron frequencies. 

The power spectrum of waves in the magnetized plasma is 
very poorly known: we assume an {\it a priori} spectrum in the form
$W(k)=W_0 (k/k_0)^{-q}$. The physical meaning underlying this
functional form is that energy is initially injected on a large
scale turbulence of size $L_c\simeq 2\pi/k_0$ and 
a cascade to smaller scales (larger values of $k$) is produced.
The fraction $\xi_A$ of the total magnetic energy $U_B$ 
in the form of waves is given by
$\xi_A=\frac{2}{U_B} \int_{k_0}^{\infty} dk_\parallel W(k_\parallel)$,
where the factor $2$ comes from the condition $W(k_\parallel)=
W(-k_\parallel)$.

The resonant interaction of particles with the electromagnetic
field of the waves results in a random walk in momentum space,
which can be described by a diffusion
coefficient $D(p)$ containing the details of the 
particle-wave interaction. Assuming that the
waves are isotropically distributed it is possible to average over
the pitch angle distribution and calculate 
the pitch-angle-averaged diffusion coefficient
(Roberts (1995)\markcite{miller} and 
Dermer, Miller and Li (1996)\markcite{dermer}):
\begin{equation}
D(p)=\frac{\pi}{2} \left[\frac{q-1}{q(q+2)}\right]
c k_0 \beta_A^2 \xi_A (r_B k_0)^{q-2} p^q/\beta.
\label{eq:diff}
\end{equation}
Here $r_B=m_e c^2 / eB$ and $\beta c$ is the particle speed.
The rate of systematic energy gain is
\begin{equation}
\frac{d\gamma}{dt} = \frac{1}{p^2} \frac{d}{dp} \left(\beta p^2
D(p)\right),
\label{eq:dedt}
\end{equation}
and the time scale for acceleration is readily given by
$\tau_{acc}=(\gamma-1)/(d\gamma/dt)$.
In the presence of energy losses with a typical time scale $\tau_{loss}$,
efficient acceleration occurs when 
$\tau_{acc}$ and $\tau_{loss}$ are comparable.  

\section{Thermalization and Acceleration}

In clusters of galaxies no stationary equilibrium
can be achieved due to the confinement of particles up to very
high energies (Berezinsky, Blasi \& Ptuskin, 1997)\markcite{bbp}. In these
conditions all the energy injected in the cluster in the form of
plasma waves and eventually converted into accelerated electrons
remains stored in the cluster, and partially dissipated only 
for Lorentz factors around $300-1000$ due to
ICS and synchrotron.
As a consequence, a time-dependent self-consistent treatment of
the acceleration and losses is required.

In the region of electron
energies that we are interested in, the acceleration process must
strongly compete with Coulomb scattering.
We develop here a suitable framework for this analysis in the
context of the Fokker-Planck equation in energy space, which allows
to take properly into account both the acceleration processes, the
processes responsible for the gas thermalization (mainly Coulomb
scattering) and the radiative losses.

The Fokker-Planck equation in energy space reads
\begin{equation}
\frac{\partial f(E,t)}{\partial t}=
\frac{1}{2}\frac{\partial^2}{\partial E^2} \left[{\cal D}(f(E,t),E)f(E,t)
\right] - \frac{\partial}{\partial E}\left[A(f(E,t),E) f(E,t)\right],
\label{eq:FP}
\end{equation}
where $E$ is the electron kinetic energy in units of $m_e c^2$, and
$f(E,t)$ is the electron distribution. $f$ is defined such that
$\int_0^{\infty} dE f(E,t)$ represents the total number of electrons
in the system, that remains constant in the absence of source terms.
The coefficients $A(f(E,t),E)$ and ${\cal D}(f(E,t),E)$ include 
all the processes
of acceleration and losses and need to be properly defined. In
general we can write ${\cal D}=D_{acc}+D_{loss}$ and
$A=A_{acc}+A_{loss}$, where $D_{acc}$ and $A_{acc}$ are the
acceleration terms while $D_{loss}$ and $A_{loss}$
account for the Coulomb energy exchange and radiative energy losses.
We discuss them separately below:

{\it i)  Stochastic Acceleration}

The diffusion coefficient in energy space is related to the diffusion
coefficient in momentum space [eq. (\ref{eq:diff})] by $D_{acc}(E)=
2\beta^2 D(p)$. The coefficient $A_{acc}$ represents the rate
of systematic energy gain and is given by eq. (\ref{eq:dedt}).

{\it ii) Thermalization and Losses}

The Fokker-Planck approach to the thermalization of an astrophysical
plasma was considered in a general case by 
Nayakshin \& Melia (1998)\markcite{melia}. 

The coefficient $A_{loss}$ for Coulomb scattering
can be written as $A_{loss}^{Coul}(f(E,t),E)=\int dE' a(E,E') f(E',t)$,
where
\begin{equation}
a(E,E')=\frac{2\pi r_e^2 c ln \Lambda}{\beta ' E'^2\beta E^2}
(E-E')\chi(E,E')
\end{equation}
and  $\chi(E,E')=\int_{E^-}^{E^+} dx x^2/
\left[(x+1)(x-1)^3\right]^{1/2}$.
Here $E^{\pm}=EE'(1\pm \beta\beta ')$, $r_e$ is the classical
electron radius, $ln \Lambda\approx 30$, and $\beta '$ ($\beta$) is the
dimensionless speed of an electron with energy $E'$ ($E$).
The diffusion coefficient
$D_{loss}^{Coul}$ associated with the Coulomb losses describes
the dispersion around the average rate of energy transport given
by $A_{loss}^{Coul}$ and is given by $D_{loss}^{Coul}(f(E,t),E)=
\int dE' d(E,E') f(E',t)$, with
\begin{equation}
d(E,E')=\frac{2\pi r_e^2 c ~ln \Lambda}{\beta \beta ' E^2 E'^2}
\left[ \zeta(E,E') - \frac{1}{2} (E-E')^2 \chi(E,E')\right]
\end{equation}
where
$$
\zeta(E,E')=\int_{E^-}^{E^+} dx \frac{x^2}{(x^2-1)^{1/2}}
\left[\frac{(E+E')^2}{2(1+x)}-1\right].
$$
(In the previous expressions some typos in the paper of
Nayakshin \& Melia (1998)\markcite{melia} have been corrected).

Note that both $A(f(E,t),E)$ and $D(f(E,t),E)$ 
depend on the function
$f(E,t)$, so that eq. (\ref{eq:FP}) is now a non-linear partial
differential equation, to be solved numerically.

When and if electrons are accelerated to high Lorentz factors the ICS
and synchrotron energy losses become important. They are included
in the Fokker-Planck equation through a coefficient $A_{loss}^{syn+ICS}$
given by the well known rate of energy losses 
(e.g. Ensslin et al. 1999\markcite{ens1}), 
while the related diffusion coefficient is neglected, since we are not
interested in the dispersion at these energies.

The Fokker-Planck equation, eq. (\ref{eq:FP}), was numerically solved
to give the electron distribution at different times. The details
of the numerical technique used to solve the equation will be given
in a forthcoming paper. We checked that, assuming no acceleration,
starting from an arbitrary distribution of electrons, Coulomb
scattering drives the system towards
a Maxwell-Boltzmann distribution with the appropriate temperature.

\section{Stochastic Acceleration in clusters of galaxies}

Stochastic acceleration due to plasma waves is a natural way to produce 
non-thermal tails in otherwise thermal particle distributions (see for
instance the applications to solar flares by Miller and Roberts 1995).
In this section we apply the formalism described above in order to calculate
the electron distribution in the intracluster medium and the related X-ray
emission. For simplicity we assume that the intracluster medium is 
homogeneous with a mean density $n_{gas}$ and a mean magnetic field $B$. 

We assume here specific values of the parameters that can explain the hard 
X-ray observations in the Coma cluster (Fusco-Femiano et al. 1998), but
this choice is not unique and a thorough investigation of the
parameter space will be carried out elsewhere. Assuming an 
emission volume with size $\sim 1.5$ Mpc, we take $n_{gas}=4\times 10^{-4}
{\rm cm}^{-3}$ and $B=0.8~\mu G$ (note that with this value of $B$ an
ICS interpretation of the hard X-rays would not be possible without 
overproducing the radio flux). 

The turbulent energy is assumed to be injected on a typical large scale
comparable with the size of a galaxy ($L_c\approx 10$ kpc)
and decay down according with a Kolmogorov spectrum ($q=5/3$). The fraction
of magnetic energy in the form of waves is taken to be $\xi_A\approx 7\%$.
Clearly, large values of $B$ and $\xi$ decrease the time scale for
acceleration. However a too efficient acceleration also means an 
efficient damping of the waves. In other words, a fast acceleration
can deplete the spectrum of the waves unless a constant injection of
energy is provided, and this injection rate is in turn constrained by the 
physical processes that we believe are responsible for the production
of waves. For instance during a merger event a total energy of the order
of $L_{mer}\sim 2\times 10^{47}$ erg/s is made available during a 
time of order 
$10^9$ yrs. We shall consider as reasonable rates of injection of energy in
the form of waves ($L_W$), values which are appreciably smaller that $L_{mer}$.
It can be shown that with the values of the parameters reported above, 
$L_W\sim 2\times 10^{46}$ erg/s.

The comparison of the time scales for Coulomb losses and acceleration
implies that the stochastic energy gain becomes efficient for $p\gtrsim 0.5$
(comparable with the minimum momentum at which waves can couple with 
electrons).
We start assuming that the initial ($t=0$) electron distribution is 
a Maxwell-Boltzmann distribution with temperature $T\approx 7.5$ keV,
and we evolve the system under the action of stochastic acceleration
and energy losses.
The results of the calculation are shown in Fig. 1 for $t=0$ (thin solid
line), $t=5\times 10^8$ yrs (thick dashed line) and $t=10^{9}$ yrs 
(thick solid line). The thin dash-dotted line is a thermal distribution
with temperature $8.21$ keV estimated to give the best fit to the thermal
X-ray emission from Coma (Hughes et al. 1993). It is clear that the energy
injected in waves is partially reprocesses into thermal energy, because of the 
efficient Coulomb scattering. Therefore the average temperature of the 
gas increases, as it is expected for instance in merger events. In addition
however a pronounced non-thermal tail appears. For $E\gg 0.1$ the non-thermal
tail has an approximately power law behaviour with a power index $\sim 2.5$
up to $E\sim 1000$ where ICS and synchrotron losses cut off the spectrum. As a
consequence, stochastically accelerated electrons do not have any effect
on radio observations.

The rate of production of X-rays per unit volume due to bremsstrahlung
emission can be calculated as
\begin{equation}
j_X(E_X,t)=n_{gas}c \int dp f(p,t)\beta \sigma_B(p,E_X)
\end{equation}
where $f(p,t)=\beta f(E,t)$ and $\sigma_B(p,E_X)$ is the differential cross
section for the production of a photon of energy $E_X$ from bremsstrahlung
of an electron with momentum $p$ (Haug 1997\markcite{haug}). 
Since we are assuming a constant density 
in the cluster, the total X-ray flux from the fiducial volume $V$ is 
$I_X(E_X)=V~j_X(E_X)/(4\pi d_L^2)$, where $d_L$ is the distance to the cluster.
We specialize these calculations to the case of the Coma cluster, assuming
$h=0.6$ for the dimensionless Hubble constant. The results are plotted 
in Fig. 2, with the lines labelled as in Fig. 1. The data points are 
from BeppoSAX while the upper limits are from OSSE (Rephaeli et al. 1994).
In this calculation we assumed that the injection of waves is continuous.
If at some point the injection is turned off, the system gradually
thermalizes and the electrons approach a Maxwell-Boltzmann distribution
in $\sim 10^7-10^8$ years. This implies that in clusters where an X-ray
tail at suprathermal energies is observed the process of wave production
must be still at work or must have been turned off not longer than 
$\sim 10^7-10^8$ years ago.

\section{Discussion and Conclusions}

We studied the thermalization process of the intracluster medium
under the effect of stochastic acceleration induced by plasma waves. 
We demonstrated that the wave-particle interactions can heat up the
intracluster gas, but the overall electron distribution is not  
exactl a Maxwell-Boltzman distribution, being characterized 
by a prominent non-thermal tail starting at the energies where 
the acceleration time is short enough to prevent a complete thermalization.
The heating process is cumulative because of particle confinement in 
clusters of galaxies (Berezinsky, Blasi \& Ptuskin (1997)\markcite{bbp}). 
Although the model was applied here to the case of the Coma cluster, 
our results are very general and the deviations from thermality are
only limited by the rate of energy injection in clusters and from the
specific cluster parameters (e.g. $n_{gas}$ and $B$). In particular,
the gas density and temperature determine the fraction of the injected
energy that is rapidly reprocessed by Coulomb scattering into thermal
energy of the bulk of electrons: high density clusters (or high density
regions in clusters) are more unlikely to have non-thermal tails.

Using the modified electron distribution in the calculation of the X-ray 
emission from clusters of galaxies results in non-thermal tails similar
to the ones recently detected by the Beppo SAX satellite. 
In the case of the Coma cluster, we showed that 
the observed spectrum can be fit reasonably well by our model, using
average values of the magnetic field which are close to the ones 
obtained from Faraday rotation measurements.

There are some observational consequences of this model, that can 
be used as diagnostic tools as more detailed observations are being 
performed. First, the efficiency of the stochastic acceleration 
depends on the ability of the plasma to recycle the injected energy 
in the form of thermal energy, through Coulomb collisions. In the 
cluster's core, where the density is larger the thermalization is more
efficient and it is correspondingly harder to form non-thermal tails
as compared with the more peripherical regions of the cluster. Thus
one prediction of the model is that the non-thermal X-ray excess would
become more prominent in the outer regions. This seems to be actually the
case for the cluster A2199 (Kaastra et al. (1999)\markcite{kaastra}) where
it was possible to measure the hard X-ray excess as a function of the
distance from the center of the cluster. 

A second consequence of the model presented here is that the modified 
electron distribution should contribute in a peculiar way to the 
Sunyaev-Zeldovich (SZ) effect: the pressure in the tail
increases the rate of up-scatter of the photons in the 
low frequency part of the photon distribution in comparison with the 
purely thermal case. This results in an enhanced
temperature change and in a different null point respect to the
thermal case (Blasi, Olinto \& Stebbins 1999\markcite{bos}). 
The SZ test is, in our
opinion the most effective in discriminating between a ICS origin and 
a pseudo-thermal origin for the hard X-ray tails in clusters of
galaxies.

\acknowledgments
I am grateful to A. Olinto, R. Rosner, C. Litwin, A. Ferrari, A. Stebbins
and A. Malagoli for many helpful discussions. I am also indebted to
S. Nayakshin, F. Melia and C. Dermer for a useful correspondence on the problem
of thermalization of astrophysical plasmas. 
I am particularly grateful to R. Fusco-Femiano for kindly providing the 
BeppoSAX data of the X-ray emission from the Coma cluster. I am also 
grateful to the anonymous referee for the several comments that helped to
improve the present paper. This work was partially  
supported by the DOE and the NASA grant NAG 5-7092 at Fermilab, by NSF through
grant AST 94-20759  and  by the DOE  through grant DE-FG0291  ER40606 at
the  University of Chicago and by I.N.F.N.

\newpage
\figcaption[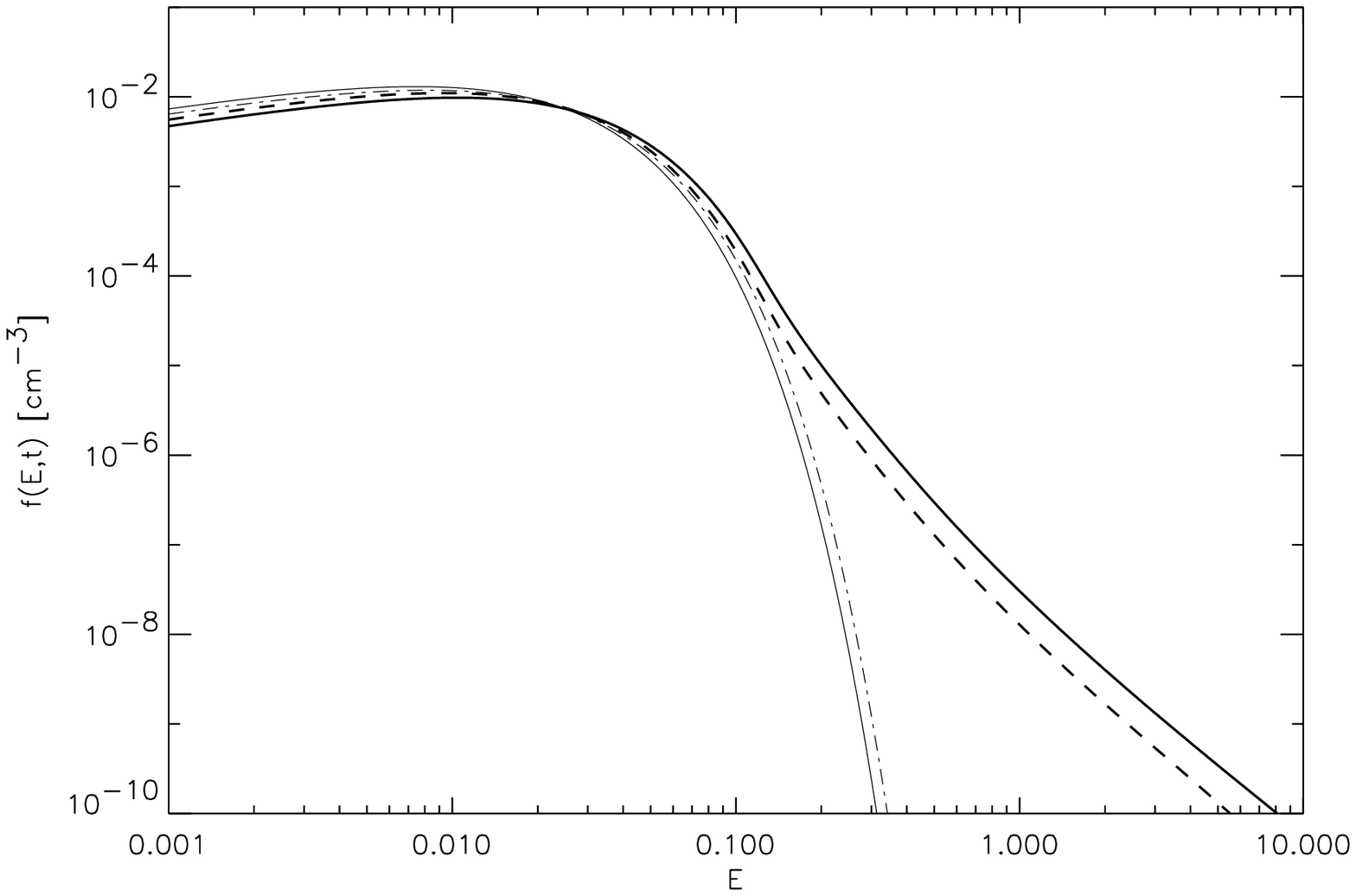]{Spectra of electrons after a time 
$t=0$ (solid thin line), $t=5\times 10^{8}$ yrs (dashed thick line)
and $10^9$ yrs (solid thick line). The dash-dot line
is a Maxwell-Boltzmann distribution with temperature $8.21$ keV.}

\figcaption[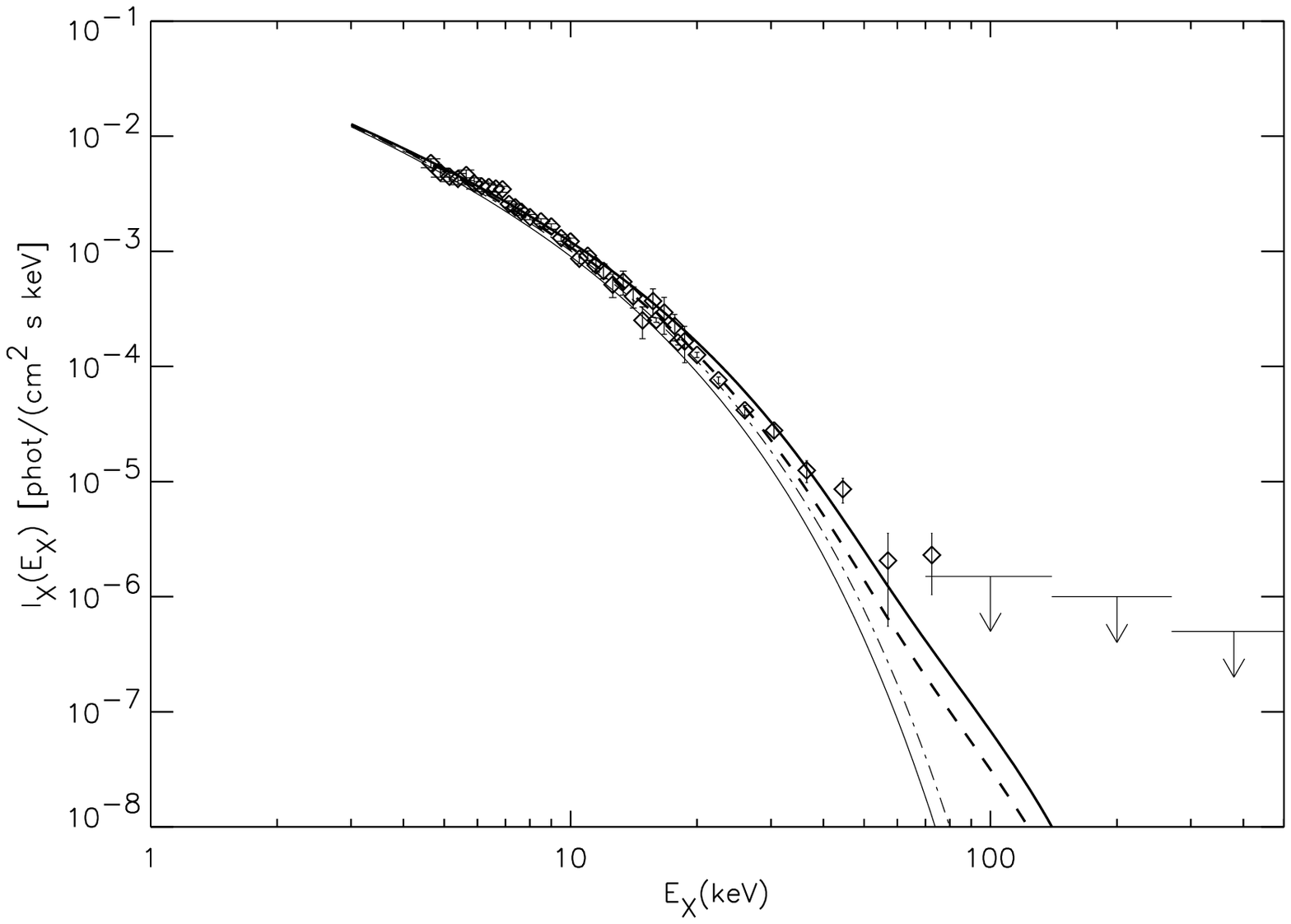]{X-ray flux from Coma as bremsstrahlung emission
of the electron distributions in Fig. 1. The data points are the result
of the BeppoSAX observations. The upper limits are from OSSE.}

\end{document}